# COMPUTATIONAL MODELING OF THE EFFECTS OF INFLAMMATORY RESPONSE AND GRANULATION TISSUE PROPERTIES ON HUMAN BONE FRACTURE HEALING

**Mohammad S. Ghiasi[1, 2], Jason E. Chen[1], Edward K. Rodriguez[3], Ashkan Vaziri[2*], Ara Nazarian[1,3,4*]**

[1]Center for Advanced Orthopaedic Studies, Beth Israel Deaconess Medical Center and Harvard Medical School, Boston, MA, USA
[2]Department of Mechanical and Industrial Engineering, Northeastern University, Boston, MA, USA
[3]Carl J. Shapiro Department of Orthopaedic Surgery, Beth Israel Deaconess Medical Center and Harvard Medical School, Boston, MA, USA
[4]Department of Orthopaedic Surgery, Yerevan State Medical University, Yerevan, Armenia

*These authors have contributed equally as senior authors.

**Corresponding Authors:**
Ashkan Vaziri, PhD
Department of Mechanical and Industrial Engineering, Northeastern University
E: vaziri@coe.neu.edu

and

Ara Nazarian, PhD
Center for Advanced Orthopaedic Studies
E: anazaria@bidmc.harvard.edu

**Abstract**

Bone healing process includes four phases: inflammatory response, soft callus formation, hard callus development, and remodeling. Mechanobiological models have been used to investigate the role of various mechanical and biological factors on the bone healing. However, the initial phase of healing, which includes the inflammatory response, the granulation tissue formation and the initial callus formation during the first few days post-fracture, are generally neglected in such studies. In this study, we developed a finite-element-based model to simulate different levels of diffusion coefficient for mesenchymal stem cell (MSC) migration, Young's modulus of granulation tissue, callus thickness and interfragmentary gap size to understand the modulatory effects of these initial phase parameters on bone healing. The results showed that faster MSC migration, stiffer granulation tissue, thicker callus and smaller interfragmentary gap enhanced healing to some extent. After a certain threshold, a state of saturation was reached for MSC migration rate, granulation tissue stiffness and callus thickness. Therefore, a parametric study was performed to verify that the callus formed at the initial phase, in agreement with experimental observations, has an ideal range of geometry and material properties to have the most efficient healing time. Findings from this paper quantified the effects of the healing initial phase on healing outcome to better understand the biological and mechanobiological mechanisms and their utilization in the design and optimization of treatment strategies. Simulation outcomes also demonstrated that for fractures, where bone segments are in close proximity, callus development is not required. This finding is consistent with the concepts of primary and secondary bone healing.

**Introduction**

Bone healing is a complex four-phase process, which starts with an inflammatory response and hematoma formation, resulting in granulation tissue development at 3-7 days post-fracture. Following this initial phase, a cartilaginous soft callus is formed from the granulation tissue in 2-4 weeks. After 2-4 months, this formation develops into a bony hard callus that surrounds the fracture site. The ossified callus is restructured for several months to years until the final bone structure is achieved, which generally resembles the original (pre-fracture) morphology of the bone [1, 2]. While the bone healing process has been experimentally studied for several decades [3-7], mechanobiological models have been used more recently to study the effects of both mechanical loading and biological factors on cellular activities and tissue formation following fracture [1, 8]. Such models can be used to study different factors that impact the healing process; to predict outcomes under different mechanical or biological conditions; and in response to new treatment strategies [9-11].

In mechanobiological modeling, mechanical factors such as strain or stress in fracture sites are typically estimated using finite element (FE) analysis. Mechanical stimuli influence biological processes and cellular activities, such as mesenchymal stem cell (MSC) migration, tissue differentiation, angiogenesis, and growth factor secretion, which in turn influence and regulate the bone healing process [1, 12-19]. Most mechanobiological models of bone healing consider a predefined callus with an ideal fixed geometry and predefined material properties [12-14, 20], where they neglect the initial phases of healing (i.e. the inflammatory response, hematoma evolution to form granulation tissue and initial callus development during the first few days post-fracture) [1]. However, few studies have accounted for callus geometry development in their simulations by assuming that it is similar to a volume expansion due to the application of thermal loading [21-23] or swelling pressure [9, 24]. However, this may not properly simulate the actual mechanism of callus geometry development, especially during the initial phase of healing [1]. Another limitation of the current studies is characterization of the material properties of the hematoma and granulation tissue during the initial phase [1, 20, 25, 26].

On the other hand, a growing body of experimental studies has highlighted the critical role of initial phases of healing on the bone healing process and outcome. For instance, inhibiting the initial post-fracture inflammatory response through anti-inflammatory treatment has been reported to impair granulation tissue formation and callus development, consequently delaying or preventing healing [27, 28]. Moreover, interfragmentary gap size and initial stability of the fracture site (i.e. fixation level of interfragmentary motion) are critical factors, which specify the form of healing (i.e. primary or secondary healing) and the recovery time. In primary bone healing, where the distance between bone fracture surfaces is very small and is completely constrained by fixation, no callus is formed. Secondary bone healing involves callus formation, where callus size partially depends on the interfragmentary motion levels conducive to healing [22, 29-33]. Moreover, the callus geometry is shown to be an optimal shape to endure the mechanical loading during the healing process. [34-36].

Therefore, *we hypothesize that the initial phase has a contributory mechanobiological effect on the overall bone healing process*, *resulting in formation of an initial callus with an ideal range of geometry and material properties to achieve the most efficient healing time*. To that end, we utilized a pre-developed finite element-based model by Lacroix & Prendergast (2002) [29] to simulate the bone healing process in models with different diffusion coefficients of MSC migration, granulation tissue Young's moduli, callus geometries, and interfragmentary gap sizes. These parameters modulate the outcome of bone healing during its initial phase, which involves inflammatory response, hematoma evolution to form granulation tissue and initial callus development during the first few days post-fracture. The diffusion coefficient can specify local levels of MSC density, especially during the initial post-fracture days [22]. The elastic modulus of granulation tissue determines the mechanical response level of the fracture site during the initial phase [1]. The mechanical response of fracture sites and MSC density depend on callus thickness during the healing process, including the initial phase [29]. Interfragmentary gap size and mechanical stability of fracture site can alter the callus thickness especially at the inflammatory response and soft callus phase [31, 37]. In this parametric study, we aim to investigate how these factors and the callus developed at the initial phase of bone healing influence healing time and healing pattern. The overall goal being to develop a simulation tool to test how early intervention, be it pharmacologic or in the form of surgical construct rigidity management, may affect the early phase of healing and result in different healing pathways as is observed clinically when direct healing without callus formation is contrasted with intramembranous ossification.

**Materials and Methods**

The mechanobiological regulation outlined by Prendergast et al. (1997) [17] was utilized to determine tissue differentiation type under applied mechanical loading (**Figure 1-A**). As a general expression, high levels of mechanical stimuli result in fibrous tissue formation, intermediate levels promote cartilaginous tissue formation, and lower levels lead to bone formation. This mechanobiological regulation was smoothed and modified based on Sapotnick and Nackenhorst's work [38], in order to prevent abrupt changes in tissue differentiation categories (**Figure 1-B**) [38].

A human bone shaft was modeled as a hollow cylinder with a transverse cut perpendicular to the cylindrical axis. An axisymmetric finite-element model of the bone was developed according to the model presented by Lacroix & Prendergast (2002) [29]. The FE model was made of 4-node quadrilateral, bilinear displacement, and bilinear pore pressure elements (**Figure 1-C**, **right**). For the base model with a 4 mm callus thickness (i.e. d=4 mm in **Figure 1-C left**) and a 3 mm interfragmentary half gap size (i.e. h=1.5 mm in **Figure 1-C left**), there were 311 elements in the marrow, 366 elements in the bone fragment and 2,034 elements in the callus (**Figure 1-C**). Boundary conditions were applied at the bottom and left borders of the model as shown in **Figure 1-C, left**. Bone, bone marrow, cartilage and fibrous tissue were modeled as linear poroelastic biphasic materials, with material properties shown in **Table-1** [20, 29]. The bone healing process was simulated for up to 120 iterations (days), with results obtained for each day using an iterative process. The iterative simulation of healing process was stopped either when 120 iterations were completed or sooner when a complete bony callus was achieved (i.e. a complete bony callus is achieved when every element of callus gains Young's modulus higher than 2 GPa). In each iteration, an axial load was applied to the top end of the bone and was increased linearly from 0 to 500 N in 1 second, similar to the model presented by Lacroix & Prendergast (2002) [29], to calculate fluid flow and octahedral shear strain for each element (ABAQUS version 6.13-2, Simulia, Providence, RI, USA). Through a separate finite element-based diffusion analysis, MSC migration was simulated to determine the spatial and temporal MSCs distribution using $\frac{\partial c}{\partial t} = D\nabla^2 c$, where $c$ is the MSC density, $D$ is the diffusion coefficient of MSC migration and $t$ is time. For the base model, value of 0.5 mm$^2$/day was considered for MSC diffusion coefficient. Bone marrow and periosteal surface of the bone and soft tissues surrounding the callus were considered as the MSC migration sources. A mesh convergence study was performed for the finite element analysis of the base model and the models with different values of MSCs diffusion coefficients to eliminate any mesh dependence in the final results. We repeated the numerical simulation for models with a wide range of diffusion coefficients of MSC migration, granulation tissue elastic moduli (denoted by *Eg*), callus thicknesses (denoted by *d*) and interfragmentary half gap sizes (denoted by *h*). To specify an appropriate range of variation for each parameter, we considered a base model [29] with normal values of 0.5 mm$^2$/day, 1 MPa, 4 mm and 1.5 mm for MSC diffusion coefficient, granulation tissue Young's modulus, callus thickness and interfragmentary half gap size, respectively. For the upper bound of MSC diffusion coefficient range, it was increased until a state of saturation was observed and for the lower bound, it was reduced until nonunion or delayed healing was observed. For other parameters, a similar approach was conducted to determine the upper and the lower bounds. However, we stopped at 2 MPa for the upper bound of granulation tissue Young's modulus, since values higher than 2 MPa are even stiffer than fibrous tissue or bone marrow, which is not probable for a relatively fresh blood clot [39]. As a result, the following domains of variables have been specified:

- [0.001, 0.01, 0.1, 0.5, 1, 10, 100] $mm^2$/day for MSC diffusion coefficient
- [0.01, 0.05, 0.1, 0.2, 0.5, 1, 2] MPa for Young's modulus of granulation tissue
- [1, 2, 3, 4, 5, 6, 7, 8] mm for callus thickness
- [0.5, 1, 1.5, 2, 2.5, 3, 3.5, 4] mm for interfragmentary half gap size.

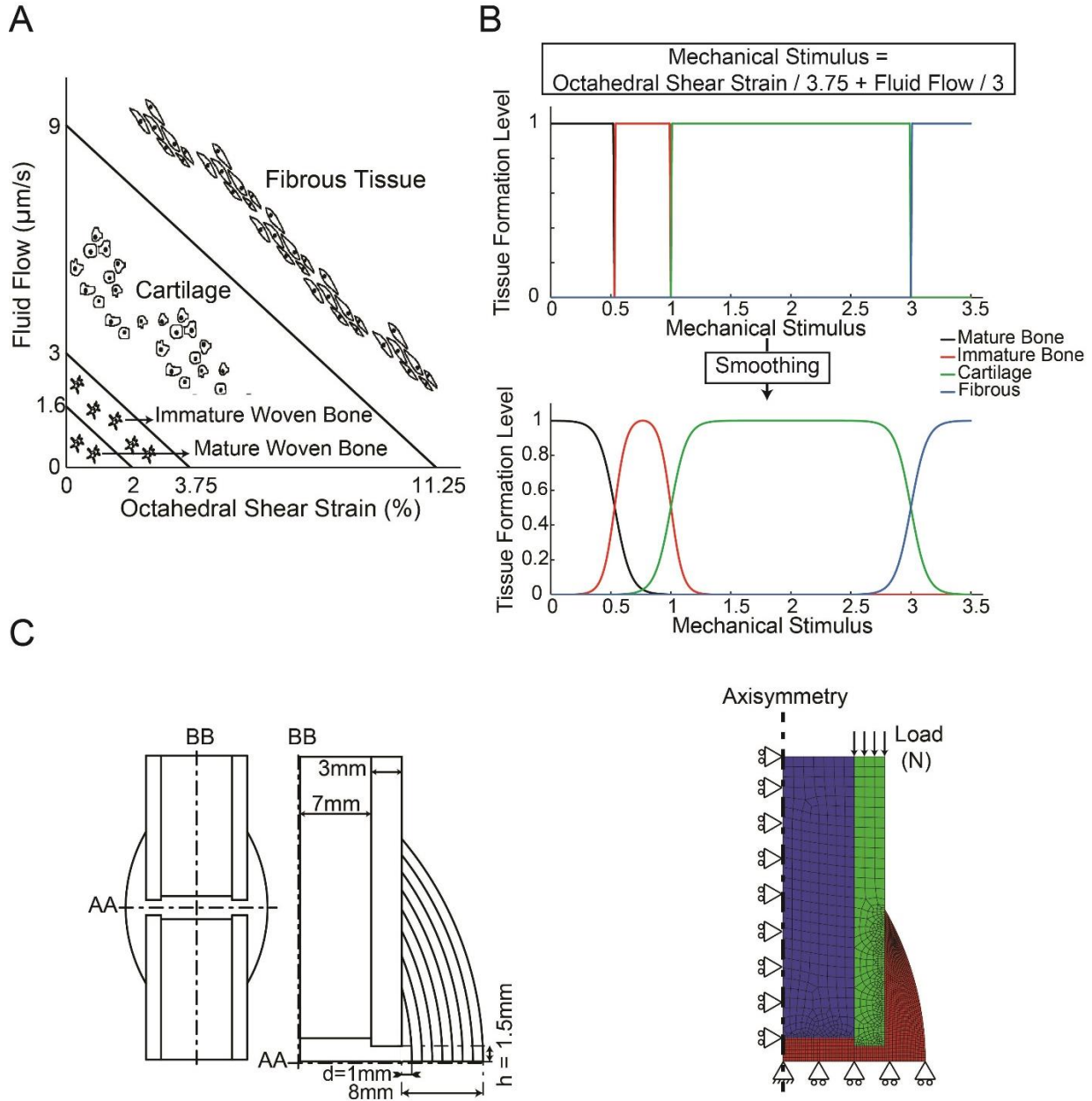

*Figure 1: A) Mechanobiological regulation by Prendergast et al. (1997) [17]. B) Smoothed mechanobiological regulation based on Sapotnick and Nackenhorst (2015) [38]. C) Left: Callus geometry dimensions including thickness (d) and interfragmentary half gap size (h). Right: FE mesh and boundary conditions of stress analysis where the blue elements are marrow, green elements are bone, and red elements are callus.*

*Table 1: Material properties*

|  | Cortical Bone | Marrow | Granulation Tissue | Fibrous Tissue | Cartilage | Immature Bone | Mature Bone |
| --- | --- | --- | --- | --- | --- | --- | --- |
| Elastic Modulus (MPa) | 20,000 | 2 | 0.001 - 2 | 2 | 10 | 1000 | 6000 |
| Permeability (m$^4$/Ns) | 1E-17 | 1E-14 | 1E-14 | 1E-14 | 5E-15 | 1E-13 | 3.7E-13 |
| Poisson's Ratio | 0.3 | 0.17 | 0.17 | 0.17 | 0.17 | 0.3 | 0.3 |
| Solid bulk modulus (MPa) | 20,000 | 2300 | 2300 | 2300 | 3400 | 20,000 | 20,000 |
| Fluid bulk modulus (MPa) | 2300 | 2300 | 2300 | 2300 | 2300 | 2300 | 2300 |
| Porosity | 0.04 | 0.8 | 0.8 | 0.8 | 0.8 | 0.8 | 0.8 |

We considered the time associated with complete development of the following structures as possible healing indices: 1) cartilaginous callus (CC), 2) bony bridging (BB), and 3) bony callus (BC) [30, 40]. It was assumed that a cartilaginous callus is developed when a cartilaginous connection is formed between two bone fragments (i.e. a sequence of elements exists with Young's modulus higher than 10 MPa to connect the bone fragment with bottom border of the callus) [40]. Bony bridging is achieved when a bony connection forms between the two bone fragments (i.e. a sequence of elements exists with Young's modulus higher than 2 GPa to connect the bone fragment with bottom border of the callus). Finally, a bony callus is achieved when every element of the whole callus has a Young's modulus greater than 2 GPa [20].

**Results**

The simulation results for models with different levels of diffusion coefficients varying from 0.001 $mm^2$/day to 100 $mm^2$/day are outlined in **Figure 2**. At the start of simulation, MSCs migrate from the three above-mentioned sources into the fracture site. For the cases with diffusion coefficient of 0.5 mm$^2$/day, level of MSCs density within the whole callus was greater than 50% of the maximum allowed cell density at day 5. When diffusion coefficient increased to 10 mm$^2$/day, level of MSCs density became greater than 50% of the maximum allowed cell density at the end of day 1. However, when diffusion coefficient decreased to 0.1 mm$^2$/day, level of MSCs was higher than 50% of the maximum allowed cell density after 40 days, and when it decreased to 0.01 mm$^2$/day, level of MSCs never reached the greater than 50% of maximum allowed cell density threshold in 120 days. In models with a small diffusion coefficient value (i.e. 0.001 to 0.01 mm$^2$/day), a long delay in healing was predicted, resulting in the formation of an incomplete bony callus after 120 days. Models with a diffusion coefficient in the range of 0.1 to 1 mm$^2$/day predicted a normal healing process with the formation of a complete bony callus within 120 days of simulation. Further increase in the diffusion coefficient affected neither the healing process nor the timeline. Moreover, interfragmentary strain reduced by 0%, 5% and 10% at day 1, day 3 and day 7, respectively, and maximum fluid flow reduced by 0%, 0% and 5% at the same days, respectively, when MSC diffusion coefficient increased from 0.5 mm$^2$/day to 100 mm$^2$/day. On the other hand, interfragmentary strain increased 0%, 7% and 50% at day 1, day 3 and day 7, respectively, and maximum fluid flow increased 0%, 0% and 20% at the same days, respectively, when MSC diffusion coefficient reduced from 0.5 mm$^2$/day to 0.01 mm$^2$/day.

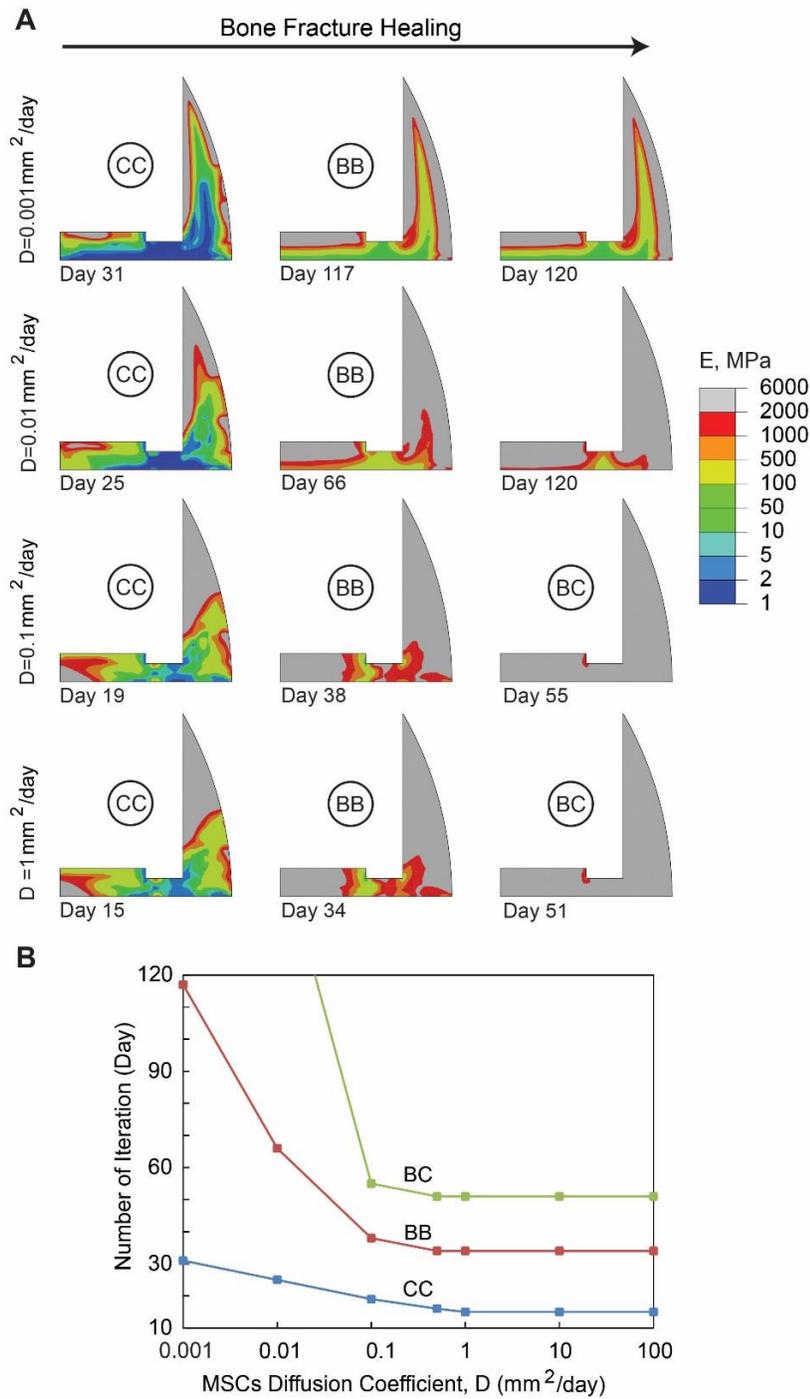

*Figure 2: A) Healing pattern at different days during the healing process. The days are selected to show the onset of cartilaginous callus (CC), bony bridging (BB) and bony callus (BC) formation in models with different diffusion coefficients, D. B) Effect of the diffusion coefficient on the healing duration (i.e. No. of days) associated with the onset of cartilaginous callus, bony bridging and bony callus formation. In this set of simulations, Eg=1 MPa, d=4 mm, and h=1.5 mm.*

The role of granulation tissue Young's modulus on the healing process is demonstrated in **Figure 3**. No considerable changes were observed in the healing outcome for elastic modulus values ranging from 0.01 to 0.2 MPa, since cartilaginous callus occurred at day 23 to 25, bony bridging occurred at day 46 to 48, and bony callus occurred at day 66 to 70. However, by increasing the elastic modulus from 0.2 MPa to 2 MPa, cartilaginous callus was formed 10 days earlier, while bony bridging occurred 16 days earlier, followed by the development of bony callus 24 days earlier. Also, interfragmentary strain reduced by 33%, 37% and 45% at day 1, day 3 and day 7, respectively, and maximum fluid flow reduced by 0%, 14% and 36% at the same days, respectively, when Young's modulus of granulation tissue increased from 1 MPa to 2 MPa. On the other hand, interfragmentary strain increased 306%, 257% and 144% at day 1, day 3 and day 7, respectively, and maximum fluid flow increased 237%, 212% and 190% at the same days, respectively, when granulation tissue Young's modulus reduced from 1 MPa to 0.1 MPa.

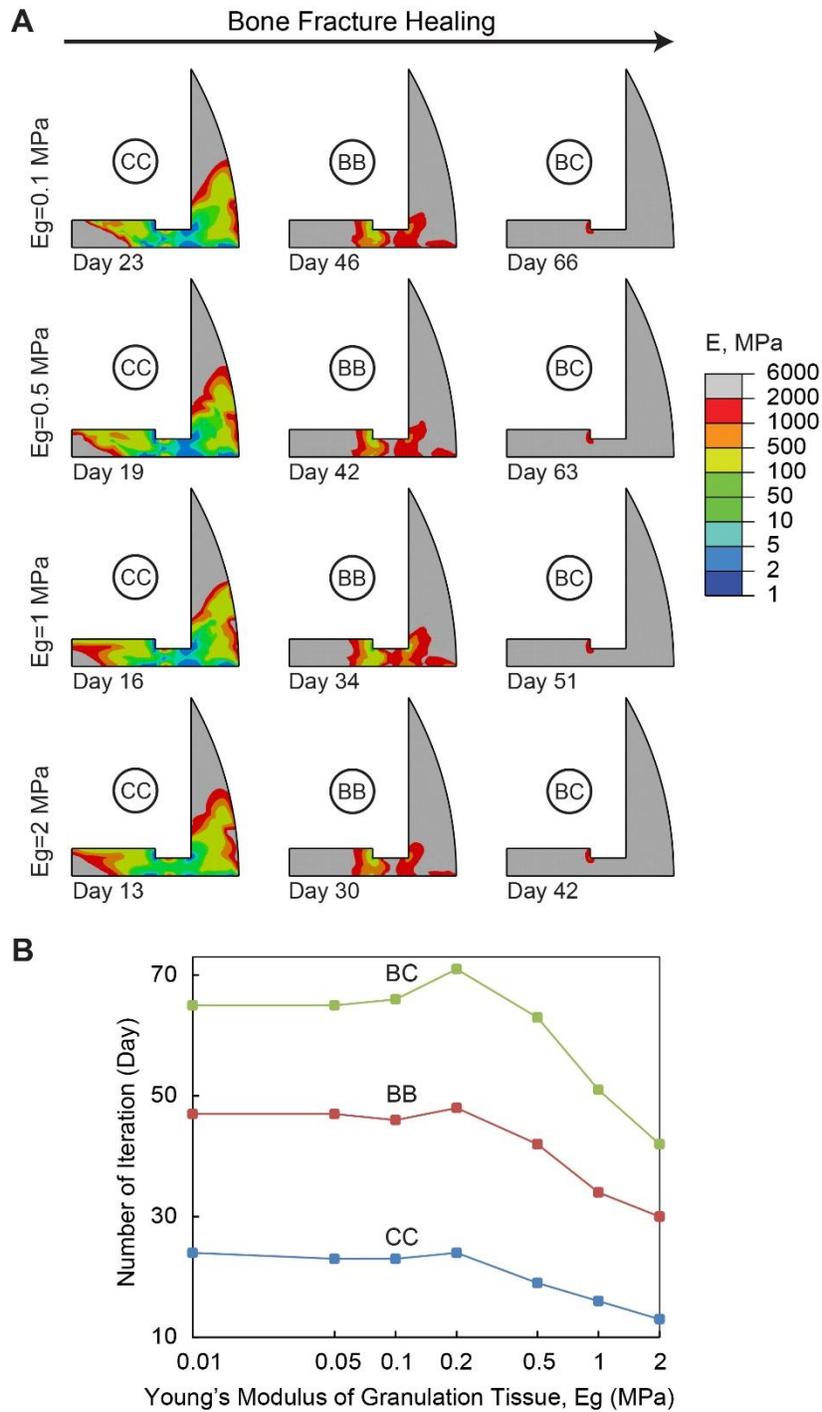

*Figure 3: A) Healing pattern at different days during the healing process. The days are selected to show the onset of cartilaginous callus (CC), bony bridging (BB) and bony callus (BC) formation in models with different elastic moduli of granulation tissue, Eg. B) Effect of granulation tissue's elastic modulus on the healing duration (i.e. No. of days) associated with the onset of cartilaginous callus, bony bridging and bony callus formation. In this set of simulations, D=0.5 mm$^2$/day, d=4 mm, and h=1.5 mm.*

The modeling results for different sizes of callus thickness are exhibited in **Figure 4**. An extremely small callus thickness (1 mm) was predicted to develop into a fibrous callus and nonunion. A small 2 mm callus thickness progressed to a cartilaginous callus in two months, a bony bridge in three months and bony callus in four months. A callus thickness range from 3 to 6 mm led to a cartilaginous callus in 2-3 weeks, bony bridge in 4-6 weeks and complete bony callus in 6-10 weeks. Callus thicknesses greater than 6 mm enhanced the speed of bone healing, as the bony callus was completed within one month for thicknesses ranging from 7-8 mm. Moreover, interfragmentary strain reduced by 3%, 31% and 75% at day 1, day 3 and day 7, respectively, and maximum fluid flow reduced by 3%, 32% and 71% at the same days, respectively, when callus thickness increased from 4 mm to 8 mm. On the other hand, interfragmentary strain increased 13%, 20% and 52% at day 1, day 3 and day 7, respectively, and maximum fluid flow increased 78%, 0% and 36% at the same days, respectively, when callus thickness decreased from 4 mm to 1 mm.

The effect of interfragmentary half gap size on bone healing, where $h$ is varied between 0.5 mm to 4 mm is shown in **Figure 5**. For a 0.5 mm interfragmentary half gap size, a cartilaginous callus was predicted at day 4, bony bridging occurred at day 13 and complete bony callus occurred in 33 days. For a 4 mm interfragmentary half gap size, cartilaginous callus was achieved in one month, bony bridging occurred in two months and complete bony callus occurred in three months. An increase in interfragmentary half gap size from 0.5 mm to 4 mm consistently delays the bone healing process, resulting in an increase in the healing time. Also, interfragmentary strain reduced by 40%, 62% and 81% at day 1, day 3 and day 7, respectively, and maximum fluid flow reduced by 0%, 25% and 59% at the same days, respectively, when interfragmentary half gap size reduced from 1.5 mm to 0.5 mm. On the other hand, interfragmentary strain increased 35%, 42% and 84% at day 1, day 3 and day 7, respectively, and maximum fluid flow increased 144%, 129% and 217% at the same days, respectively, when interfragmentary half gap size increased from 1.5 mm to 4 mm.

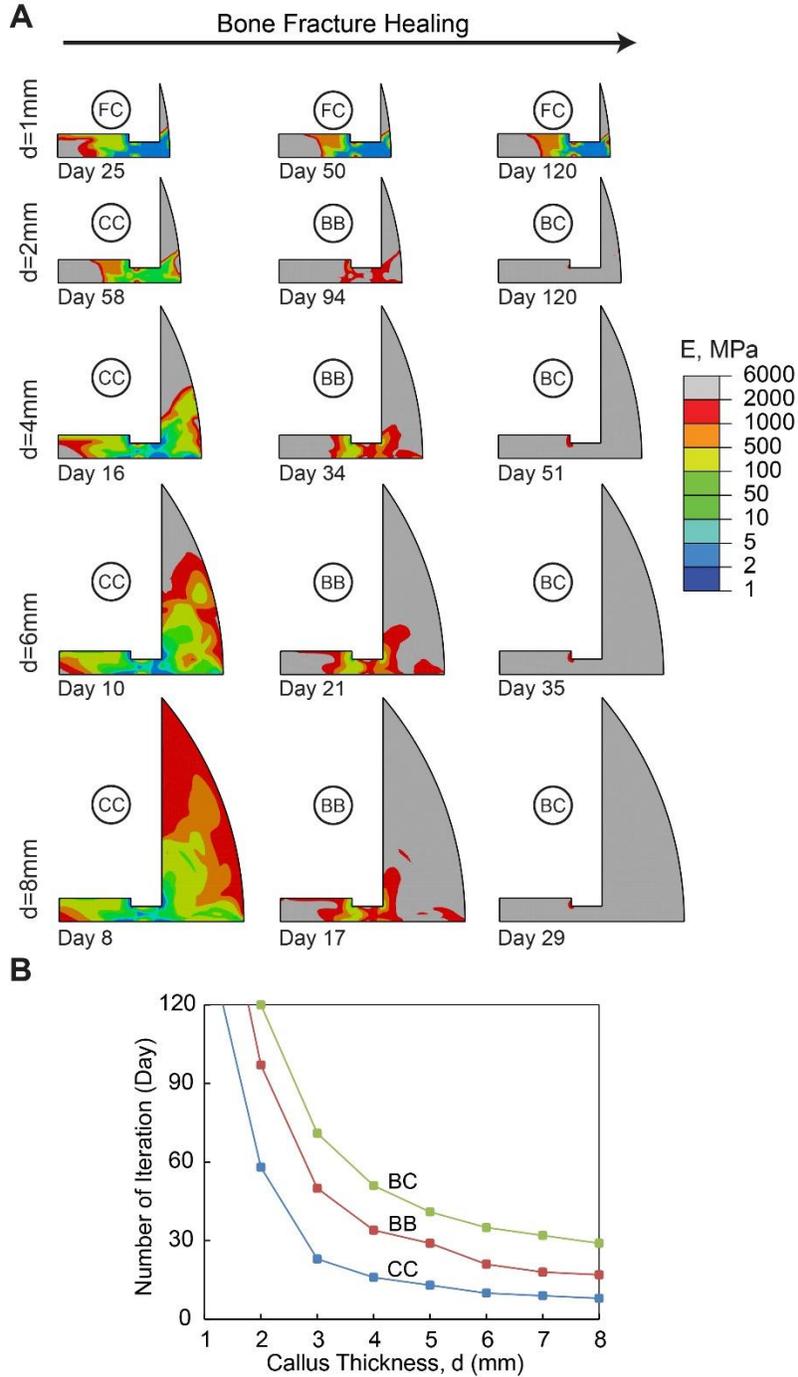

*Figure 4: A) Healing pattern at different days during the healing process. The days are selected to show the onset of cartilaginous callus (CC), bony bridging (BB) and bony callus (BC) formation in models with different callus thicknesses, d. B) Effect of callus thicknesses on the healing duration (i.e. No. of days) associated with the onset of cartilaginous callus, bony bridging and bony callus formation. In this set of simulations, D=0.5 mm2/day, Eg=1 MPa, and h=1.5 mm.*

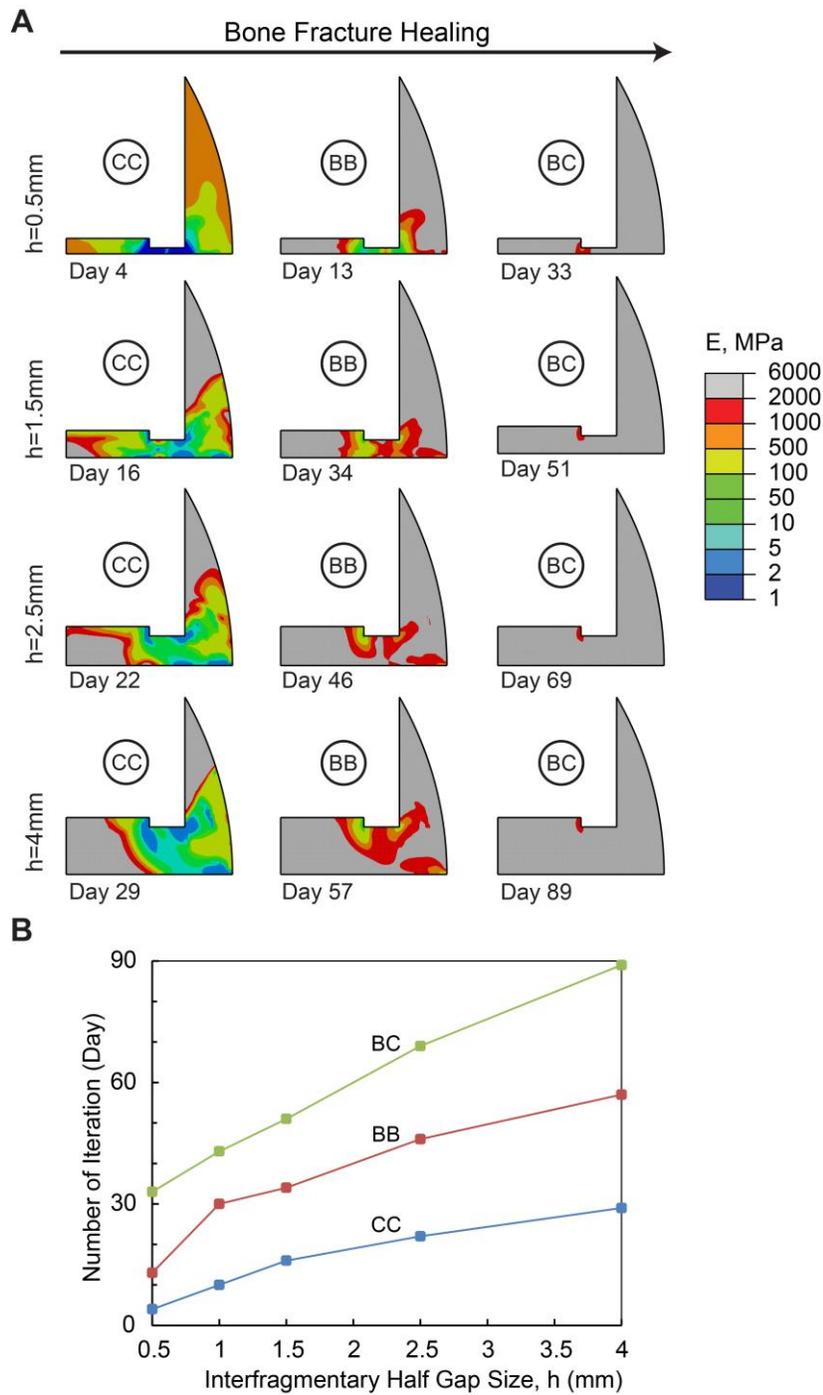

*Figure 5: A) Healing pattern at different days during the healing process. The days are selected to show the onset of cartilaginous callus (CC), bony bridging (BB) and bony callus (BC) formation in models with different interfragmentary half gap sizes, h. B) Effect of interfragmentary half gap sizes on the healing duration (i.e. No. of days) associated with the onset of cartilaginous callus, bony bridging and bony callus formation. In this set of simulations, D=0.5 mm2/day, Eg=1 MPa, and d=4 mm.*

The day corresponding to the onset of bony bridging for three different callus thicknesses ($d$=3, 5, and 7 mm) is shown in **Figure 6**, where the MSC diffusion coefficient is varied between 0.01 and 10 mm$^2$/day. The results are presented for three different values of granulation tissue Young's modulus ($Eg$=0.1, 1 and 2 MPa). It should be noted that for the callus thickness of 1 mm, boney bridging does not occur in 120 days in the simulations, regardless of the level of MSC diffusion coefficient and granulation tissue Young's modulus considered in this set of simulations. Thus, no results are shown for the callus thickness of 1mm. In general, the onset of bridging occurs quicker for the models with a thicker callus. Faster MSC migration and a stiffer granulation tissue also expedite the healing, resulting in a quicker formation of boney bridging.

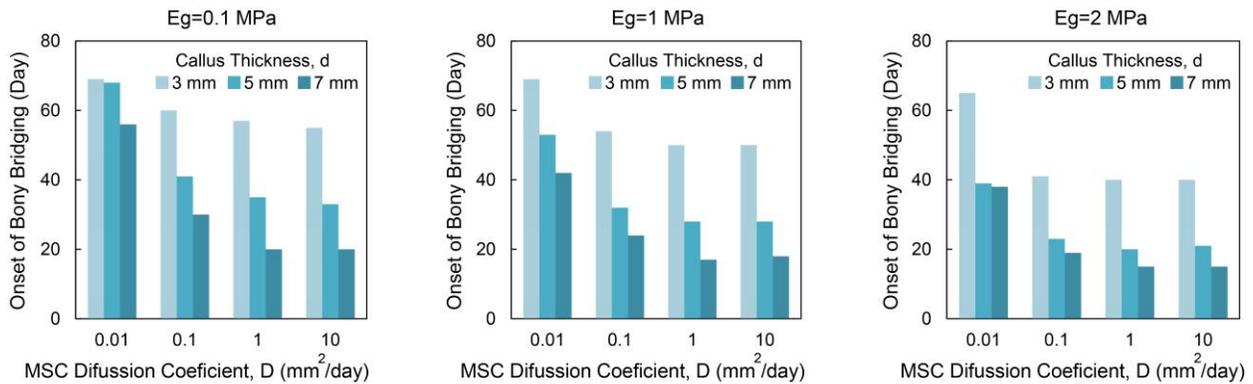

*Figure 6: Onset of bony bridging in models with different callus thicknesses, MSC diffusion coefficient and Young's modulus of granulation tissue.*

**Discussion**

We used a well-established model of the bone healing process presented by Lacroix & Prendergast (2002) [29] to design a parametric study in order to computationally quantify effects of the initial phase of healing on the healing outcome. For the base model, our numerical simulations predict that cartilaginous callus is achieved in 2-3 weeks from the start of the healing process, bony bridging occurs in one month, and complete bony callus is developed in less than 2 months. This development timeline matches fairly well with clinical observations, as well as the results presented in previous numerical investigations [5, 20, 29]. In general, models with a larger value of diffusion coefficient for MSC migration, a stiffer granulation tissue and a thicker callus thickness predict lower level of mechanical stimuli and faster healing process. An increase in the diffusion coefficient for MSC migration means that the MSCs can be distributed more rapidly across the callus area for differentiation. In our simulations, when the diffusion coefficient was less than 0.1 mm$^2$/day, our simulations indicated insufficient supply of MSCs to support differentiation and tissue formation, which subsequently delayed healing or resulted in nonunion. This is consistent with the nonunion results predicted by Geris et al. [11], when the MSC sources of migration were removed. On the other hand, no considerable change in the healing process was observed by increasing the diffusion coefficient to values greater than 1 mm$^2$/day. At this stage, MSCs are present in high volume in the callus, and thus the healing time is rather limited by MSC differentiation or tissue formation rates. In other words, MSCs are readily available throughout the callus, but no improvement in healing occurs, as MSCs cannot differentiate and form tissue at a faster rate [13, 41].

In addition, callus development serves to support mechanical loading and provide the desired stability for bone healing [8, 34, 35]. Hence, calluses with small thicknesses or those made of soft granulation tissue are not able to support the applied mechanical loading and provide a secure environment for the requisite cellular activity. Based on this study, callus thicknesses smaller than 3 mm or granulation tissue softer than 0.5 MPa resulted in delayed healing or nonunion. On the other hand, a callus thicker than 6 mm does not result in improvements in healing. Larger callus size results in prolonged MSCs migration throughout the callus, as well as resorption and remodeling [42]. Granulation tissues with an elastic modulus higher than 2 MPa are even stiffer than fibrous tissue or bone marrow, which is not probable for a relatively fresh blood clot [39]. Therefore, after a certain level, there is no need for a larger or stiffer callus to support mechanical loading and stabilize the fracture site. According to the findings of this study, there is an ideal range that has also been observed in experimental studies[5, 43-45] for each initial phase parameter (i.e. 0.1-1 mm$^2$/day for migration rate, 1-2 MPa for Young's modulus of granulation tissue, 3-6 mm for callus thickness). For instance, MSCs mostly spread out over the callus during the initial phase of healing in both our simulations with the ideal range of migration rate and experimental observations by Iwaki et al.[43], Zhang et al.[46] and Wang et al.[47]. The granulation tissue indentation modulus, measured by Leong et al.[25] at 35 days post-fracture in a rat, completely matches the ideal range of granulation tissue Young's modulus predicted in this paper. The above mentioned ideal ranges of callus geometry and gap size were also in agreement with the experimental observations made by Aguat et al.[31, 48], Claes et al.[37] Boer et al. [49], Epari et al.[50] and Yang et al.[51]. Thus, simulation results interestingly demonstrate that the formed callus at the initial phase of healing (i.e. a normal healing that is observed in experimental studies and clinical environments) contains optimal geometry and material properties to have the most efficient healing time.

As indicated by our results, increasing the interfragmentary gap size delays bone healing, and shrinking the gap expedites it [29, 37]. This was seen in simulations with a 0.5 mm interfragmentary half gap size, where bony bridging and complete bony callus formation occurred in two weeks and one month, respectively. The remarkable impact of smaller interfragmentary gap size motivated us to investigate its effects on the smallest callus sample with the thickness of 1 mm (i.e. the sample where no sign of healing was seen in 120 days when combined with a 1.5 mm interfragmentary half gap size) (**Figure 4**). Interestingly, a normal pattern of healing was observed when a very small 0.25 mm interfragmentary half gap size was paired with a very small 1 mm-thick callus (**Figure 7**). The results matched experimental and clinical observations [31, 32, 37] and emphasize that a larger callus is necessary, when the interfragmentary gap is enlarged, in order to have a normal pattern of healing. **Figure 7** also indicated that if bone fragments were tightly positioned with respect to one another in the secondary form of bone healing, almost no callus development would be needed which was in agreement with the concept of primary bone healing [4, 52]. These findings highlight the potential capability of bone healing models in understanding the basis and plausible mechanisms behind clinical observations [10].

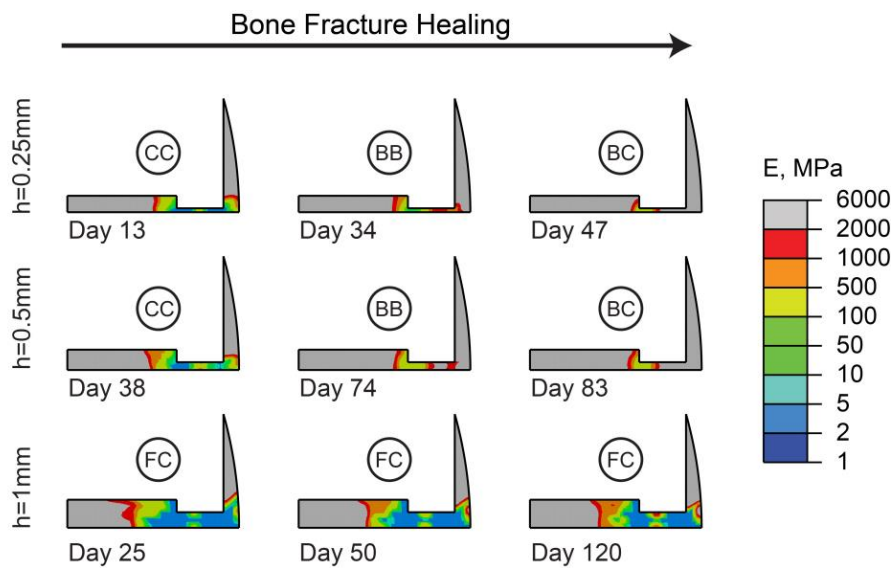

*Figure 7: Healing pattern at different days during the healing process. The days are selected to show the onset of cartilaginous callus (CC), bony bridging (BB) and bony callus (BC) formation in models with different interfragmentary half gap sizes, h. In this set of simulations, D=0.5 mm2/day, Eg=1 MPa, and d=1 mm.*

The quality of cartilaginous callus, position of bony bridging, and pattern of healing can also be affected by changes in the initial phase of healing. An increase in diffusion coefficient shifts the bony bridging position from the exterior of the callus towards the middle, and increase the average stiffness of the cartilaginous callus. Increasing the interfragmentary gap size also changes the position of bony bridging from the exterior of the callus to the inside. However, in some cases, it is not entirely clear how the initial phase affects the healing pattern. For instance, no differences were observed in the bony bridging position or cartilaginous callus stiffness, following the change in callus thickness or elastic modulus of granulation tissue.

As one of the limitations of this study, we only focused on material properties and geometrical factors of the initial callus as the outcome of the initial phase of healing. Other factors such as angiogenesis, growth factors effects, oxygen tension or type of loading were not directly investigated, since complementary experimental studies are needed to provide reliable data in order to include them in the simulation. Also, material properties of the granulation tissue in the initial phase of healing, including elastic modulus or diffusion coefficient for MSC migration, have not been studied and analyzed well under different conditions of healing [1, 20]. Therefore, a precise range of material properties is not available for the initial callus formed at the initial phase to compare with our simulation results. However, some estimates have been conducted in previous simulation studies for the material properties of granulation tissue, which are in agreement with our reported optimal range [13, 41]. Moreover, we assumed that the callus size was fixed after the initial phase of healing. This assumption is consistent with the clinical observation, where the callus geometry develops during the initial phase of healing and is resorbed during the remodeling phase [5, 31, 37].

In conclusion, we have outlined the importance of the initial phase of healing, resulting in formation of the initial callus with a range of geometry and material properties for optimal healing time. Findings from this paper quantified the effects of the four important initial phase parameters on healing outcome, which can be utilized to design and optimize treatment strategies by tuning these parameters. Consequently, there are well-established models to simulate soft callus formation, hard callus development, and remodeling phases of healing; however, one part is missing to complete the puzzle and that is the initial phase of healing. This study emphasizes that the initial phase of healing should not be ignored in modeling of the healing process. Results from this study also raise questions about the clinical applications and the mechanisms of the initial healing phase such as how can we regulate these parameters at the initial healing phase to achieve the most efficient healing time? and how do micro motions of fracture site, biological factors and immune system response influence callus size and the level of granulation tissue formation at the initial phase of healing? Hence, further experimental investigations on the biological and mechanical factors in early stage of healing are required to develop more robust and predictive models that can simulate healing from the beginning to the end, and to better understand how clinicians are able to control and modulate the initial phase with its parameters.

**Competing Interests**

The authors declare no competing interests.

**Authors' Contributions**

- Mohammad S Ghiasi conducted the computational modeling, helped with the conceptualization and wrote the manuscript.
- Ara Nazarian, Ashkan Vaziri and Edward K Rodriguez contributed in conceptualization, design of the study and review of the manuscript.
- Jason E Chen helped with analysis of the results and writing the manuscript.

**Acknowledgments**

The authors would like to acknowledge Professor Hashemi from Northeastern University for fruitful discussions on this work.

**Funding**

This work was supported by in part by a grant from the United States National Science Foundation's Civil, Mechanical, and Manufacturing Innovation (Grant No. 1634560), and in part by the Department of Mechanical and Industrial Engineering at Northeastern University.